\documentclass[12pt]{iopart}
\usepackage{graphicx}
\begin{document}

\title{Charged hadron results from Au+Au at 19.6 GeV}

\author{D. Cebra for the STAR Collaboration}
\address{Physics Department, University of California, Davis,CA, 95616, USA}
\ead{cebra@physics.ucdavis.edu}

\begin{abstract}

Results from a one day $\sqrt{s_{NN}} = 19.6$ GeV Au+Au test run at RHIC
using the STAR detector are presented. The quality of these results from 
only 175,000 triggered events demonstrates some of STAR's physics 
capabilities for the upcoming beam energy scan at RHIC. From these 19.6 GeV 
Au+Au collisions, we have analyzed the transverse mass spectra of
$\pi^{\pm}$, $K^{\pm}$, $p$, and $\bar{p}$ at midrapidity and $m_T - m_0$ 
$< 1.0$ GeV/$c^2$. We have also measured the two-pion interferometry source 
radii. The collision energy ($\sqrt{s_{NN}} = 19.6$ GeV) of this 
low energy Au+Au RHIC collider run is very close to that of the 158 AGeV 
fixed-target Pb+Pb runs at the SPS ($\sqrt{s_{NN}} = 17.3$ GeV).
We present comparisons between these STAR data and the   
results published by NA49, NA44, WA98, and CERES.

\end{abstract}



\section{\label{sec:intro}Introduction}

One of the goals of relativistic heavy-ion experiments is to 
determine the properties of nuclear matter over a wide range of 
temperatures and densities. This requires an understanding of
interactions within the medium at both the partonic and hadronic levels. 
At sufficient densities, heavy-ion collisions are expected to form
a quark-gluon plasma (QGP). The plasma expands, cools, and hadronizes at the 
chemical freeze-out point. The hadron gas continues to expand 
and cool until it reaches a point of kinetic freeze-out. Although the
single-particle spectra and the HBT radii are not fixed until the final 
stage of the collision, they still provide important constraints allowing us 
to model the evolution of nuclear matter from high energy density to kinetic
freeze-out. 

In this analysis, we study the charged hadron spectra and the HBT radii from 
Au+Au collisions at $\sqrt{s_{NN}}$ = 19.6 GeV at RHIC. 
At this collision energy, we are studying heavy-ion collisions with an 
energy density similar to that created in the top energy (158 $A$GeV) 
fixed-target Pb+Pb collisions at the CERN SPS (with a corresponding 
$\sqrt{s_{NN}}$ of 17.3 GeV). 
The results presented in this paper provide an important 
cross check between the collider program at RHIC and the fixed-target 
heavy-ion program at the SPS. These measurements provide a baseline to 
allow better interpretation of the results from the higher energy 
collisions studied at RHIC and for the proposed future low energy 
running of RHIC. 

\section{\label{sec:star}The STAR Experiment}

The data presented in this study have been obtained using the STAR 
experiment. These data were taken with a minimum bias
trigger on the final day of the
RHIC run II heavy-ion running period. 
This beam energy had not been planned prior
to closing the experimental halls and little time was available either for
trigger optimization or data taking. The total STAR data set with no
quality cuts was 175466 events. This is a small data sample by RHIC or SPS
standards; however one of the strengths of the STAR detector is the wide
coverage in both rapidity and azimuth. Therefore, even with the
modest number of events recorded, we still obtained inclusive spectra and 
HBT radii.

\begin{figure}[t]
\centering
\includegraphics[width=1.0\textwidth]{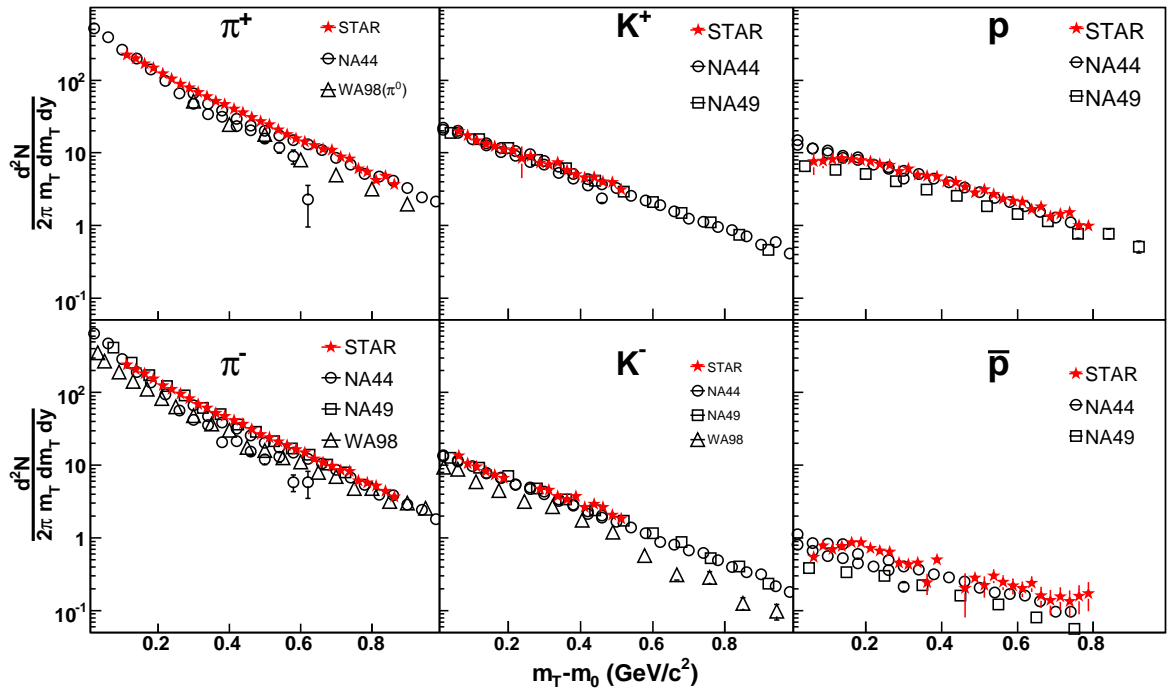}
\caption{The 19.6 GeV transverse mass spectra compared with the 
results of SPS experiments NA44, NA49, and WA98. All the results 
correspond to the most central impact parameter and rapidity data sets 
analyzed by each experiment. 
The solid red stars are the STAR 19.6 GeV results, 
the open squares NA49~\cite{NA49}, 
the open circles NA44~\cite{NA44},
and the open triangles WA98~\cite{WA98}.
All the SPS results are from the Pb+Pb runs at $\sqrt{s_{NN}} = 17.3$ 
GeV.} 
\label{multispectra_sps_compare}
\end{figure}

\section{\label{sec:results}Results}

Fig.~1 shows the transverse mass distributions of
$\pi^+$, $\pi^-$, $K^+$, $K^-$, $p$, and $\bar{p}$ at midrapidity, $-0.1 <
y < 0.1$, from the top 10\% most central Au+ Au collisions at 19.6 GeV.
These midrapidity charged particle spectra are compared with SPS Pb+Pb 
results at 17.3 GeV (NA44~\cite{NA44}, NA49~\cite{NA49},and 
WA98~\cite{WA98}). 
In general, the spectra are quite similar to those measured at
the SPS, although close attention should be paid to the differences 
in $N_{\rm{part}}$, $\sqrt{s_{NN}}$, and rapidity range. NA49 has the 
hardest centrality selection. The extra 2.3 GeV of center of mass energy 
increases the midradipity yields of all particle, except protons. STAR has 
the strictest radipity definition. The NA44 and WA98 experiments have 
$m_t-m_0$ dependent rapidity acceptances, which effect the shapes of the 
spectra. 

\begin{figure}
\centering  
\includegraphics[width=0.45\textwidth]{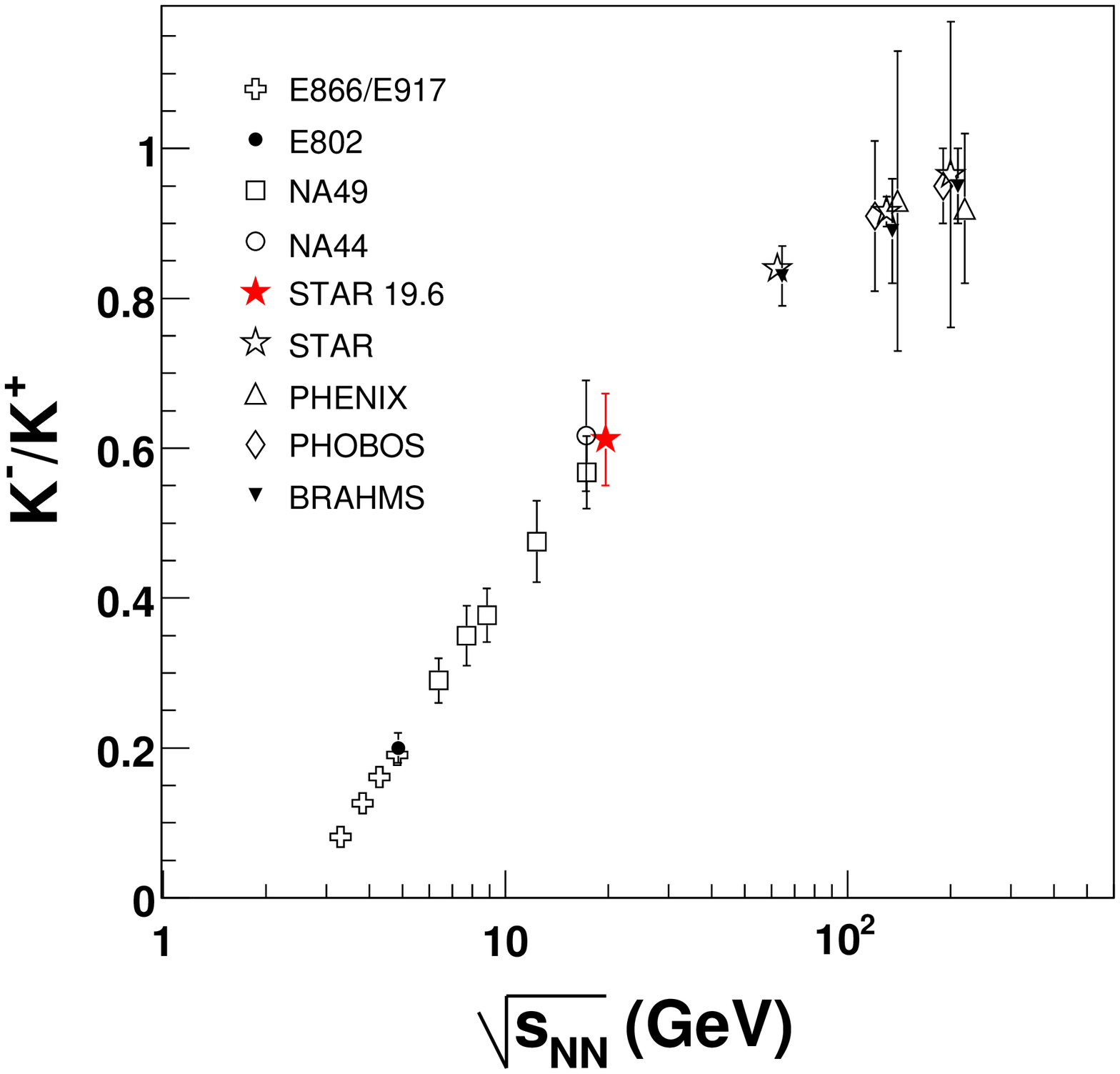}
\includegraphics[width=0.50\textwidth]{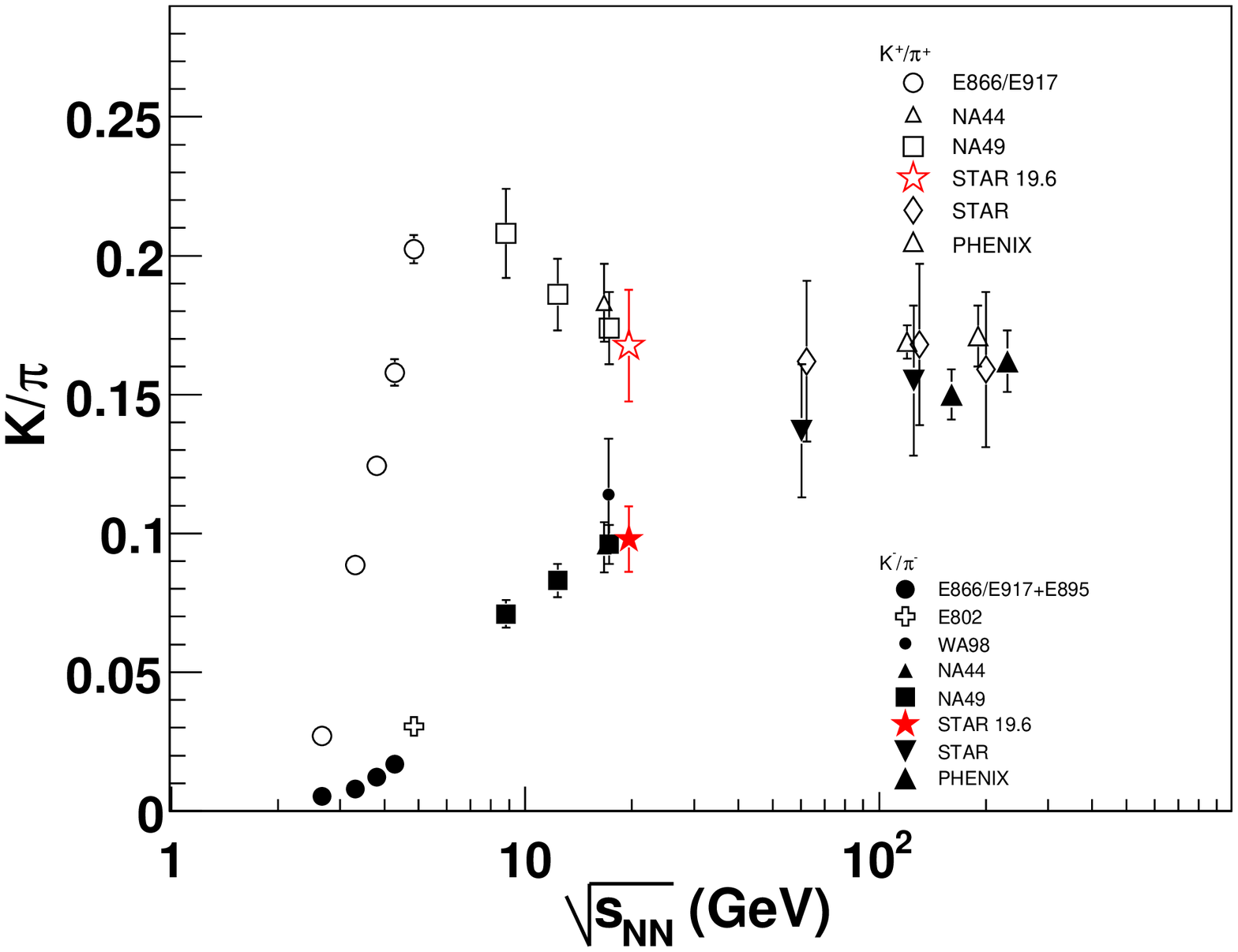}
\caption{The $K^-/K^+$ and $K/\pi$ ratios as a function of 
$\sqrt{s_{NN}}$ for the most central collisions at midrapidity.
The published values from E866/917~\cite{E866}, 
E802~\cite{E802}, NA49~\cite{NA49}, NA44~\cite{NA44}, WA98~\cite{WA98},
PHENIX~\cite{PHENIX}, PHOBOS~\cite{Phobos}, BRAMHS~\cite{BRAHMS},
and STAR~\cite{STAR} are shown for comparison.}
\end{figure}

In Fig.~2,  we compare the most central midrapidity
$K^-/K^+$ and $K/\pi$ ratios with other
experiments (E866/917~\cite{E866}, E802~\cite{E802}, NA44~\cite{NA44},
NA49~\cite{NA44}, WA98~\cite{WA98}, PHENIX~\cite{PHENIX}, 
Phobos~\cite{Phobos}, BRAHAMS~\cite{BRAHMS}, and STAR~\cite{STAR}). 
The 19.6 GeV STAR data agree well with the trends established by the
previously published data. As the energy increases, the $K^-/K^+$ ratio 
rises toward unity. The $p\bar/p$ ratio is found to be $0.10 \pm 0.01$, 
which is consistent with SPS results. 
In a QGP, the energy threshold for producing an $s\bar{s}$ pair is lower
than in a hadron gas. To study strangeness production, we look at the
ratios of charged kaons, which carry the bulk of produced strangeness, and
pions, the most abundantly produced hadrons from the collisions. 
The $K^+/\pi^+$ and $K^-/\pi^-$ ratios are shown in
Fig.~2. The excitation functions of the positive and negative ratios are 
found to vastly differ which may be suggestive about the nature of the 
medium. 

We have also studied the two-pion interferometry as a function of $m_t$.
Fig. 3 shows a direct comparison with STAR's measurement of Au+Au collisions 
at $\sqrt{s_{NN}}=19.6$ GeV and Pb+Pb results at $\sqrt{s_{NN}}=17.3$ 
GeV from CERES~\cite{CERESHBT} and NA49~\cite{NA49HBT}. As seen, STAR 
results are consistent with those from the SPS experiments.

\begin{figure}
\centering
\includegraphics[width=0.45\textwidth]{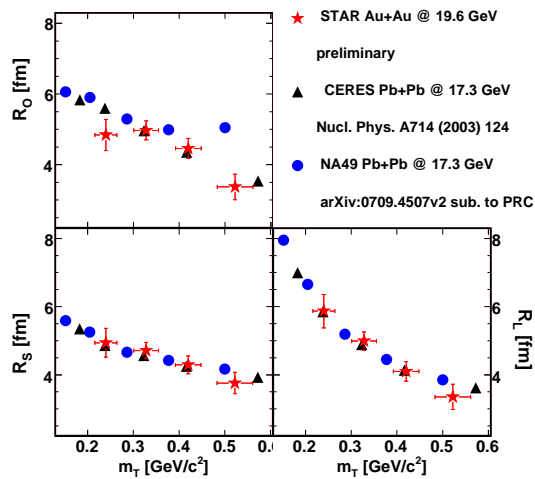}
\caption{The two-pion interferometry radii as a function of $m_t$ compared 
to the results of CERES~\cite{CERESHBT} and NA49~\cite{NA49HBT}.}
\end{figure}

\section{\label{sec:summary}Summary}

In summary, STAR has measured the transverse mass spectra 
for $\pi^\pm$, $K^\pm$, $p$, and $\bar{p}$ from Au+Au
collisions at $\sqrt{s_{NN}} = 19.6$ GeV. Comparisons of spectra
and particle ratios are made to the published SPS and RHIC 
results. After accounting for differences between number of 
participants, the rapidity of the data, and the bombarding energy, 
the yields were comparable for all experiments. 
The $K^-/K^+$ ratio is about 0.6 and $\bar{p}/p$ about 0.1. The $K/\pi$ 
ratios are consistent with the established trends as a function of 
$\sqrt{s_{NN}}$. The pion interferometry results are consistent with the 
published SPS results.

\end{document}